\documentclass[twocolumn,showpacs,amsmath,amssymb,aps,prd,floats]{revtex4-1}
\usepackage{graphicx}
\usepackage{dcolumn}
\usepackage{bm}
\usepackage{epsfig}
\usepackage{xspace}

\def\apj{The Astrophysical Journal}
\def\apjl{The Astrophysical Journal Letter}
\def\apjs{The Astrophysical Journal Supplement}

\def\aap{Astronomy and Astrophysics}

\def\nat{Nature}
\def\mnras{Mon. Not. R. Astron. Soc.}

\begin{document}

\title{Constraints on the intergalactic magnetic field  from ${\gamma}$-ray observations of GRB 190114C}
\author{Ze-Rui Wang$^{1,2}$, Shao-Qiang Xi$^{1,2}$, Ruo-Yu Liu$^{1,2}$, Rui Xue$^{1,2}$, Xiang-Yu Wang$^{1,2}$}
\affiliation{$^1$School of Astronomy and Space Science, Nanjing University, Nanjing, 210093, China;\\
$^2$Key laboratory of Modern Astronomy and Astrophysics (Nanjing University), Ministry of Education, Nanjing 210023, People's Republic of China}

\begin{abstract}
Very high energy photons from cosmological gamma-ray bursts (GRBs) are expected to interact with extragalactic background light (EBL) and produce electron-positron pairs when they propagate through intergalactic medium (IGM). These relativistic pairs will then up-scatter cosmic microwave background (CMB) photons and emit secondary GeV emission. Meanwhile, the  motion of these pairs are deflected by intergalactic magnetic field (IGMF), so the secondary GeV photons arrive later than the primary emission. It has been suggested that the properties  of the secondary GeV emission can be used to constrain IGMF. Recently,  TeV gamma-ray emission has been detected, for the first time, from a GRB (GRB 190114C) by the MAGIC telescope and its steep ${\rm \gamma-ray}$ spectrum shows a clear evidence of absorption by EBL. We then constrain the IGMF with the GeV flux limit obtained from the $Fermi$-LAT observations. We find a limit of $>10^{-19.5}$ G for the coherence length of $\lambda \leq 1$ Mpc. Although this limit is weaker than that obtained by using blazars, it represents the first limit from ${\rm \gamma-ray}$ observations of GRBs, which provides  an independent constraint on IGMF. We also find that, for transient ${\rm \gamma-ray}$ sources, one can choose a favorable time window to search for the echo emission at a particular energy.
\end{abstract}

\maketitle

\section{Introduction}
As the weakest magnetic field, the intergalactic magnetic field (IGMF) exists in the voids of large-scale structure in the universe \citep{2013A&ARv..21...62D}. According to the dynamo theory, IGMF can be the seed field where the magnetic fields in galaxies and galaxy clusters are created \citep{2008RPPh...71d6901K}. Detection and measurement of the properties of IGMF are important to assess the dynamo theory in galaxies and galaxy clusters. In addition, the origin of IGMF is largely unknown. There are two classes of models for the seed fields: 1) astrophysical models, which assume that the seed fields are generated by motions of the plasma in galaxies, and 2) cosmological models, in which the seed fields are produced in the early universe. The properties of IGMF will be also useful to constrain its origin.

TeV $\gamma$-ray sources can be used to constrain the properties of IGMF \citep{2013A&ARv..21...62D, 2013A&A...554A..31N}. Very high energy photons interact with EBL and produce electron-positron pairs when they propagate through IGM. The relativistic pairs will then upscatter CMB photons and emit secondary GeV emission through IC radiation. Meanwhile, the directions of these pairs are deflected by IGMF. So, these secondary GeV photons will reach the observer with different directions from that of the primary emission, which is called ``pair halo'', and different times, which is called ``pair echo''. These differences will influence the observation properties of secondary GeV photons. It has been suggested that the properties (duration and strength) of these GeV photons can be used to constrain the IGMF.

For persistent sources (e.g., TeV blazars), the pair halo method is more suitable for constraining IGMF  \citep{1994ApJ...423L...5A, 2010MNRAS.406L..70T, 2011MNRAS.414.3566T, 2010ApJ...719L.130N, 2011ApJ...733L..21D, 2014ApJ...796...18A, 2015ApJ...814...20F}. {Ref.~\citep{2007A&A...469..857D} suggested to combine (non)-observation of blazars by Fermi and IACT as a tool to
probe the IGMF.} Ref.~\citep{2010Sci...328...73N} obtained a limit of $B>3 \times 10^{-16}$ G for a coherence length $\lambda = 1$ Mpc. It is difficult to get a better limit on IGMF from blazars, since the persistent primary GeV emission will pollute the cascade emission. To remove the interference by the primary GeV emission, Ref.~\citep{2011MNRAS.414.3566T} considers TeV blazars without GeV emission, and found a lower limit of $10^{-15}$ G for IGMF. Besides, in Ref.~\citep{2018ApJS..237...32A}, the authors suggest the limit on IGMF is larger than $3 \times 10^{-16}$ G for coherence length $\lambda \gtrsim 10^{-2}$ Mpc by using both spectral and spatial information.

The pair echo method can also be used to constrain IGMF \citep{1995Natur.374..430P, 2002ApJ...580L...7D, 2004A&A...415..483F, 2008ApJ...686L..67M}. In Ref.~\citep{ 2012ApJ...744L...7T}, they obtained a limit of $>10^{-20}{\rm G}$ for IGMF  by using the quasi-simultaneously observed GeV-TeV light curves of Mrk 501. GRBs, as being predicted to be TeV sources, have also been proposed to be useful in  constraining IGMF through the pair echo method. As transient sources, most of their echo photons will arrive at the observer at a later time than that of the primary photons, therefore one can easily distinguish those photons and obtain the limit on IGMF. Since no GRBs with TeV emission were detected previously, all studies on limiting IGMF assumed a theoretically expected TeV flux \citep{2004ApJ...604..306W, 2004ApJ...613.1072R, 2008ApJ...682..127I, 2009MNRAS.396.1825M, 2011MNRAS.410.2741T}. For example, assuming that the GeV spectrum of GRB 13027A extending to multi-TeV band \citep{2017ApJ...847...39V}, the authors suggest that the limit on IGMF is larger than $3 \times 10^{-17}$ G for a coherence length $\lambda = 1$ Mpc.

Recently, TeV emission has been detected, for the first time, from a GRB (GRB 190114C) by the MAGIC telescope \citep{2019Natur.575..455M,2019Natur.575..459M}, which makes it possible to study the pair echo of the TeV $\gamma$-rays from GRBs. This burst has a redshift of $z = 0.42$ \citep{2019GCN.23695....1S}. The observed spectrum of the TeV emission of GRB 190114C shows a clear steepening caused by the EBL absorption. In this paper, we will give {constraints} on IGMF by studying the possible pair echo emission of GRB 190114C. The method for calculating the echo emission flux is shown in section~\ref{sec:echo}, and the results for constraining  IGMF are shown in section~\ref{sec:obs}. We give discussions and conclusions  in section~\ref{sec:sec5}. The cosmological parameters of $H_{0}= 68$ km $\rm s^{-1} Mpc^{-1}$, $\Omega_{\Lambda} = 0.7$ and $\Omega_{\rm M} = 0.3$ are used in the following calculation.

\section{The Echo Emission} \label{sec:echo}

\subsection{Flux of the echo emission}\label{sec:exact}

The cascade process occurs when the TeV photons are absorbed by EBL during the propagation through IGM.  The distribution of pairs $P(\gamma_{\rm e},\epsilon)$ produced in interaction between TeV photons with {dimensionless energy $\epsilon={\rm h}\nu/{{\rm m_e}{\rm c}^2}$} and soft photons with distribution $n_{0}(\epsilon_{0})$ is given by \citep{1988ApJ...335..786Z}

\begin{equation} \label{eq:1}
\begin{split}
P(\gamma_{\rm e}, \epsilon)=\int_{{\epsilon}/{x_{\gamma}}}^{\infty}&{\rm d} \epsilon_{0}n_{0}(\epsilon_{0})\frac{3\sigma_{\rm T}{\rm c}}{4\epsilon^2\epsilon_{0}}[r-(2+r)\frac{\epsilon}{\epsilon_{0}x_{\gamma}}\\
&+2(\frac{\epsilon}{\epsilon_{0}x_{\gamma}})^2+2\frac{\epsilon}{\epsilon_{0}x_{\gamma}}\rm ln\frac{\epsilon_{0}x_{\gamma}}{\epsilon}],
\end{split}
\end{equation}
where ${\rm x}_{\gamma}=4\gamma_{\rm e}\gamma_{\rm e}'$, $r=(\gamma_{\rm e}/\gamma_{\rm e}'+\gamma_{\rm e}'/\gamma_{\rm e})/2$, $\gamma_{\rm e}'=\epsilon-\gamma_{\rm e}$.


For a GRB with a differential luminosity $L_{\rm GRB}(\epsilon)$ and a duration timescale $t_{\rm GRB}^{\rm obs}$ for the TeV emission, the number of absorbed TeV photons $n_{\rm TeV}(\epsilon)$ can be calculated by
\begin{equation} \label{eq:3}
n_{\rm TeV}(\epsilon)=\frac{L_{\rm GRB}(\epsilon)\Delta t_{\rm GRB}}{\epsilon{\rm m}_{\rm e}{\rm c}^2}\{1-{\rm exp}[-\tau_{\gamma\gamma}(\epsilon)]\},
\end{equation}
where  $\tau_{\gamma\gamma}(\epsilon) $is the pair production optical depth due to the EBL absorption.
We derive the pairs energy spectrum as
\begin{equation} \label{eq:2}
\frac{{\rm d}N_{\rm e}(\epsilon)}{\rm d\gamma_{\rm e}}=n_{\rm TeV}p(\gamma_{\rm e}, \epsilon),
\end{equation}
where $p(\gamma_{\rm e}, \epsilon)= 2P(\gamma_{\rm e}, \epsilon)/\int{{\rm d}\gamma_{\rm e}P(\gamma_{\rm e}, \epsilon)}$ is the normalized pair distribution.

The duration of the echo emission depends on the energy of the TeV photons, the energy of the resultant  pairs, and the magnetic field, so  we define it as $t_{\rm dur}(\epsilon,\gamma_{\rm e},B)$.
The average echo flux at frequency ${\nu}$ during an observation time $t_{\rm obs}$  is given by
\begin{equation} \label{eq:4}
\begin{split}
F_{\nu}=&\frac{m_{\rm e}c^{2}}{4\pi D^{2}h}\int {\rm d}\epsilon_{0}n_{0}(\epsilon_{0})\int {\rm d}\gamma_{\rm e} C(\epsilon_{\rm echo}, \gamma_{\rm e}, \epsilon_{0})\\
&t_{\rm IC(\gamma_{\rm e})}\int {\rm d}\epsilon \frac{{\rm d}N_{\rm e}(\epsilon)}{{\rm d}\gamma_{\rm e}}\frac{1}{{\rm max}(t_{\rm dur}(\epsilon,\gamma_{\rm e},B),t_{\rm obs})},
\end{split}
\end{equation}
where $C(\epsilon_{\rm echo}, \gamma_{\rm e}, \epsilon_{0})$ is the Compton kernel \citep{1968PhRv..167.1159J}, $t_{\rm IC}(\gamma_{\rm e})$ is the cooling timescale of relativistic pairs of energy $\gamma_{\rm e}$ due to inverse-Compton (IC) scatterings. Since the IGMF is very weak, the synchrotron radiation can be ignored, and the pairs lose energy mainly through IC radiation.

The duration of the echo emission is mainly determined by the deflection angle of the pairs caused by IGMF. The defection angle depends on the coherence length $\lambda$ of the magnetic field and the distance $l_{\rm IC}(\gamma_{\rm e})$ that pairs lose energy. If $\lambda>l_{\rm IC}(\gamma_{\rm e})$, the motion of pairs can be approximated by the motion in a homogeneous magnetic field, so the angle of pairs deflected in IGMF is $\theta_{\rm B}(\gamma_{\rm e},B)=l_{\rm IC}(\gamma_{\rm e})/R_{\rm L}(\gamma_{\rm e},B)$ \citep{2008ApJ...682..127I}, where $R_{\rm L}(\gamma_{\rm e},B)$ is the Larmor radius. If the coherence length $\lambda$ is less than $l_{\rm IC}(\gamma_{\rm e})$, the deflecting angle $\theta_{\rm B}(\gamma,B)$ is modified by a factor of $\sqrt{\lambda/{l}_{\rm IC}(\gamma_{\rm e})}$.  The duration of the echo emission is $t_{\rm B}(\epsilon,\gamma_{\rm e},B)\simeq (l_{\gamma\gamma}(\epsilon)+l_{\rm IC}(\gamma_{\rm e}))\theta_{\rm B}^{2}(\gamma_{\rm e},B)/(2c)$  \citep{2008ApJ...682..127I},  where  $l_{\gamma\gamma}(\epsilon)$ is the mean free path of TeV photons, which is related with the source distance $D$ by $l_{\gamma\gamma}(\epsilon)=D/\tau_{\gamma\gamma}(\epsilon)$. In addition, due to the beaming effect, an observer sees up to beaming angle of $\gamma_{\rm e}^{-1}$ from the line of sight, so there is an angular spreading time $t_{\rm A}(\epsilon,\gamma_{\rm e})\approx (l_{\gamma\gamma}(\epsilon)+l_{\rm IC}(\gamma_{\rm e}))/(2\gamma_{\rm e}^{2}c)$. The duration of the echo emission should be the maximum of three timescales, i.e.,  $t_{\rm dur}(\epsilon,\gamma_{\rm e},B)={\rm max}(t_{\rm B}(\epsilon,\gamma_{\rm e}, {\rm B}), t_{\rm A}(\epsilon,\gamma_{\rm e}), t_{\rm GRB}^{\rm obs})$, where $t_{\rm GRB}^{\rm obs}$ is the duration of the TeV emission of GRBs. {From the MeV-GeV data, the power-law decay of the TeV emission of GRB 190114C is inferred to start from 6 s to 2454 s after the burst, so we take $t_{\rm GRB}^{\rm obs} = 2448$ s \citep{2019A&A...626A..12R, 2019ApJ...884..117W}}. In most cases, $\gamma_{\rm e}^{-1}$ is much smaller than $\theta_{\rm B}$ and $l_{\rm IC}(\gamma_{\rm e})$ is much smaller than $l_{\gamma\gamma}(\epsilon)$, so the duration time is dominated by $l_{\gamma\gamma}(\epsilon)\theta_{\rm B}^{2}(\gamma_{\rm e},B)/(2c)$.

\begin{figure}[htbp]
\centering
\includegraphics[width=1\columnwidth]{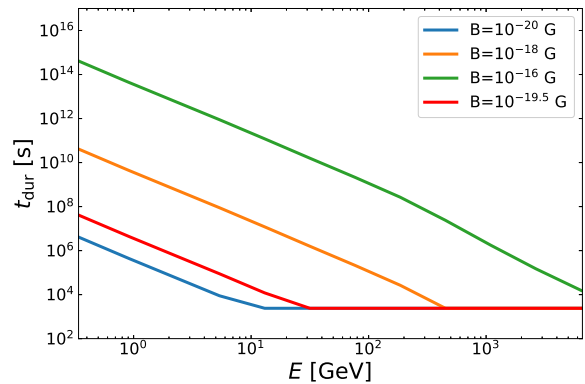}
\caption{The duration of the echo emission as a function of the photon energy for various values of the strength of IGMF.  The coherence length of IGMF is  assumed to be $\lambda = 1$ Mpc. \label{fig:durationtime}}
\end{figure}

\subsection{A crude estimate of the echo emission fluence and duration}\label{sec:rough}

Section~\ref{sec:exact} presents an accurate calculation method of the echo emission flux. In this section, we give a crude, but more straightforward estimate of the echo emission fluence, which may be useful to  guide the use of the accurate calculation.
The maximum fluence of the echo emission can be estimated from $\int S_{\nu}^{\rm max}{\rm d}\nu = E_{\rm GRB}^{\rm abs}/(4\pi D^{2})$, where $E_{\rm GRB}^{\rm abs}$ is the energy of the absorbed TeV photons. The expected fluence of the echo emission is related with the observation time and duration by
\begin{equation}\label{eq:5}
S_{\nu}\propto\left\{
\begin{aligned}
\frac{t_{\rm obs}}{t_{\rm dur}}S_{\nu}^{\rm max},& & t_{\rm obs}\leq  t_{\rm dur}~\\
S_{\nu}^{\rm max} & , & t_{\rm obs}>t_{\rm dur}.
\end{aligned}
\right.
\end{equation}

We can also obtain a rough estimate of the duration of the echo emission at different energy, assuming that the energy of pairs is half of the energy of TeV photons, i.e., $\gamma_{\rm e} = \epsilon /2$. Then the typical energy of echo photons is $\epsilon_{\rm echo}=(4/3)\gamma_e^2 \epsilon_{\rm CMB}$, and we can rewrite those two timescales as  $t_{\rm B}(\epsilon_{\rm echo}, B)$ and $t_{\rm A}(\epsilon_{\rm echo})$, respectively.  The results of relationship between the duration  and photon energy are shown in Figure~\ref{fig:durationtime}.
As the magnetic field strength increases or the photon energy decreases, the duration  becomes longer. Motivated by this, we suggest that, to obtain the best limit on IGMF, one can choose a favorable time window to search for the echo emission at a particular energy.  This will be discussed in the following section.
\begin{figure}[htbp]
\centering
\includegraphics[width=1\columnwidth]{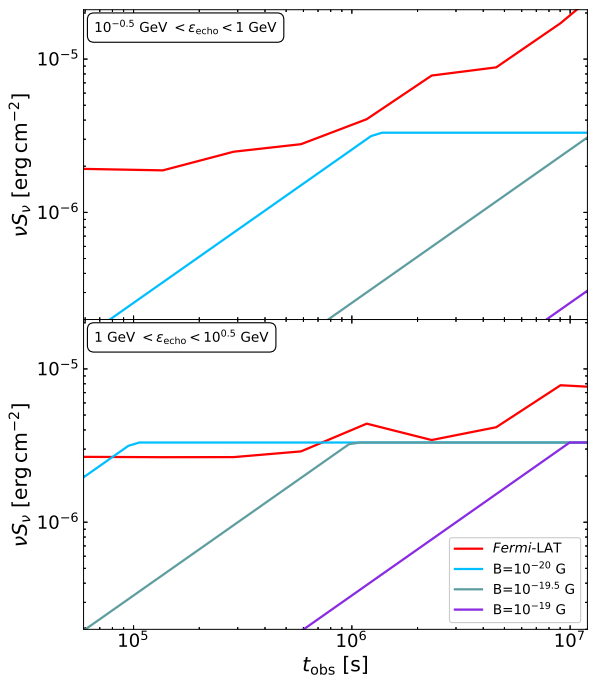}
\caption{The fluence of the echo emission as a function of the observation time for the energy bin of 0.3-1 GeV (top) and 1-3 GeV (bottom), respectively. The red lines show the fluence limit imposed by $Fermi$-LAT. Other  lines represent the crude estimates of the echo emission fluence using the method in Section~\ref{sec:rough}. The coherence length of the IGMF is assumed to be $\lambda = 1$ Mpc.  \label{fig:result1}}
\end{figure}

\section{Limits on IGMF using GRB 190114C} \label{sec:obs}
GRB 190114C has GeV emission up to $\sim$ 10000 s since the burst trigger time $T{\rm 0}$, which can be explained by the GRB prompt and/or afterglow emission \citep{2019Natur.575..459M}. We expect the echo emission time at GeV band longer than $\sim 10^5$ s assuming a magnetic field larger than $10^{-20}$ G. Thus, we first search for the possible echo emission in the $Fermi$-LAT data starting from $T{\rm 0} + 20000$ s, which is selected  to exclude  the impact of primary GeV emission, to $T{\rm 0}+9$ months, which is limited by the observation times for this GRB. We select all the source class events detected by the $Fermi$-LAT between 100 MeV and 100 GeV in a circular radius of interesting (ROI) of $12^{\circ}$ centered at the position of ($\alpha_{J2000} = 54.503^{\circ}$,$\delta_{J2000} = -26.938^{\circ}$). In order to reduce the contamination from Earth limb, all events with zenith angle $< 90^{\circ}$ are filtered out. We employ an unbinned maximum likelihood technique to test the presence of echo emission using the likelihood-ratio test statistic (TS), which is defined as twice the logarithm of the maximum of the likelihood value for the alternative hypothesis ($L_1$) divided by that for the null hypothesis($L_0$): ${\rm TS = 2(log}L_1-{\rm log}L_0)$. The null hypothesis for the test is represented by a baseline model including all point sources from the fourth LAT source catalog \citep{2019arXiv190210045T} within a circular ROI enlarged by $5^\circ$, as well as the Galactic and isotropic diffuse emission templates provided by the Fermi-LAT Collaboration\footnote{\textit{gll\_iem\_v07.fits} and \textit{iso\_P8R3\_SOURCE\_V2\_1.txt}; https://fermi.gsfc.nasa.gov/ssc/data/access/lat/BackgroundModels.html}. The spectral normalization of each source is left free to vary. The alternative hypothesis is represented by the baseline model plus a new point source at the position of GRB 190114C  located by the Swift-BAT observation\footnote{https://gcn.gsfc.nasa.gov/gcn3/23724.gcn3}, which is modeled by a power-law spectrum with free index and normalization. We find TS value is $\sim 1$, implying that  there is no detection of the echo emission. We further search for the echo emission choosing different time windows (i.e., 1 days, 15 days, 1 month, 3 months, 6 months after $T{\rm 0} + 20000$ s) at 6 logarithmically spaced energy window in $\rm 100\ MeV - 100\ GeV$. There are also no significant detections and the upper limit fluxes are then derived at a $95\%$ confident level.


In Figure~\ref{fig:result1}, we show the fluence limit (the red lines) of the echo emission  as a function of time. The fluence limit is nearly a constant for a short observation time, while it goes as $S_{\nu}^{\rm limit}\propto t_{\rm obs}^{1/2}$ when the observation is long enough.
We calculate the expected fluence  of the echo emission using the data of GRB 190114C \citep{2019Natur.575..455M}.  {The total energy radiated in TeV emission during the period from 6 s to 2454 s after the burst is $E_{0.3-1{\rm TeV}}\approx 2\times 10^{52}$ erg}.  The photon index of the intrinsic spectrum is  $-2.22^{+0.23}_{-0.25}$ and we use $-2$ in the calculation for simplicity.

\begin{figure}[htbp]
\centering
\includegraphics[width=1\columnwidth]{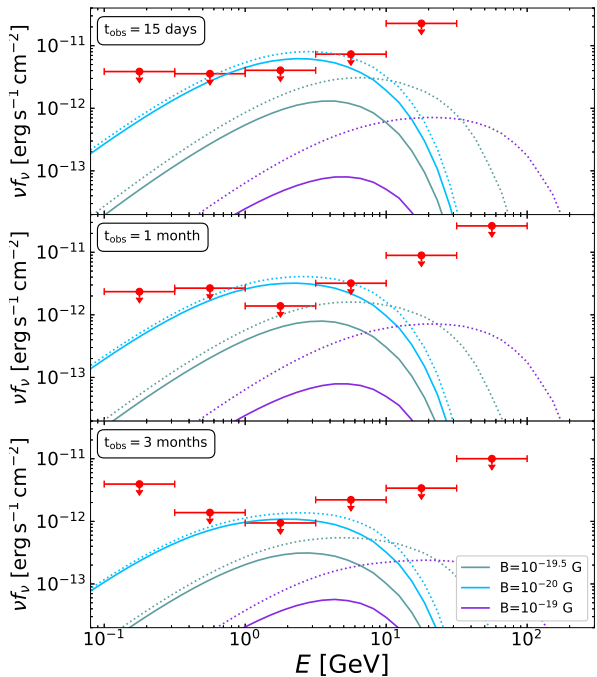}
\caption{The  spectral energy distribution of the echo emission averaged over the observation time $t_{\rm obs}$, i.e., 15 days (top), 1 month (middle) and 3 months (bottom) after $T0+20000$ s.
The red points denote the $Fermi$-LAT upper limit fluxes.
The solid lines and dotted lines represent the echo emission spectrum for primary photons with a maximum  energy of 1 TeV and 15 TeV, respectively.
\label{fig:result2}}
\end{figure}
The energy of the absorbed photons in 0.3-1 TeV will be reprocessed into that of echo photons in the energy range of 0.3-3 GeV, considering the distribution function of Eq.~\ref{eq:1}.
Then, we can estimate the fluence of echo emission in the energy range of 0.3-1 GeV and 1-3 GeV respectively. It can be found in Figure~\ref{fig:result1} that the fluence reaches its maximum value later for a stronger IGMF. The lower limit of IGMF can be estimated by comparing the theoretical fluence with the fluence limit given by $Fermi$-LAT. The case ($10^{-20}~{\rm G}\leq B < 10^{-19.5}$~{\rm G}) can be ruled out since their theoretical fluence is larger than the fluence limit. In addition, we found that the most favorable time window for constraining IGMF is about 1 month, since at this time the limit on IGMF is the best.

To obtain a more accurate result, we select three time windows close to $t_{\rm obs} = 1~{\rm month}$ and  calculate the echo emission flux using  Eq.~\ref{eq:4}. The result is shown in Figure~\ref{fig:result2}. The best constraint on IGMF comes from the case $t_{\rm obs} = 1~{\rm month}$, and longer or shorter observation times both give worse constraints on IGMF. This is consistent with the above rough estimate using the fluence. We also calculate the case where the maximum energy of TeV photons reaches 15 TeV (dotted lines in Figure~\ref{fig:result2}), and find that it does not significantly improve the limit on IGMF.
Note that  we have removed the contribution from the high energy echo photons if they arrive before  the search time window (i.e.,  $t_{\rm dur}(E) < 20000$~s). Since the duration $t_{\rm dur}(E)$ of the echo emission depends on the strength of IGMF,  the spectrum of the echo emission at the high energy end also depends on the strength of IGMF, as shown in Figure~\ref{fig:result2}.
\begin{figure}[htbp]
\centering
\includegraphics[width=1\columnwidth]{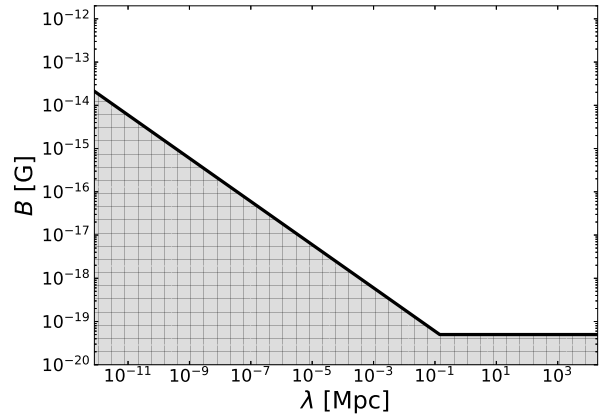}
\caption{ Observational bounds on the strength and correlation length of IGMF. The white area is the allowed parameter space. \label{fig:result3}}
\end{figure}

The coherence length $\lambda$ affects the constraints on the strength of IGMF. The lower limit on $\lambda$ is set by the requirement that the resistive magnetic diffusion time scale
has to be larger than the age of Universe, whereas the upper limit is set only by the size of the visible part of the Universe \citep{2010Sci...328...73N}.
By searching the magnetic field strength from $10^{-20}$ G to $10^{-16}$ G and coherence length from $10^{-12}$ Mpc to $10^{4}$ Mpc, we calculate the limit on IGMF for different $\lambda$. The white area in Figure~\ref{fig:result3} shows the allowed parameter space for IGMF in our case. The limit on the IGMF is $B \lambda^{1/2}=1.89\times 10^{-20}~{\rm G}~{\rm Mpc}^{1/2}$ for $\lambda<0.1$ Mpc.

\section{Discussions and Conclusions}\label{sec:sec5}

As the first  GRB with TeV emission being detected, GRB 190114C is used to constrain  IGMF. We find that  the best limit on IGMF can be derived when the observation time matches the  duration of the echo emission.   The main assumptions adopted in our calculation are as follows:
1) The effect of the second generation pairs are not considered. The optical depth due to EBL absorption is $\tau_{\gamma\gamma}(\epsilon)>1$  for the photons with energy greater than $200$ GeV at a distance corresponding to $z=0.42$. Since the flux of the echo emission with energy greater than $200$ GeV is quite low in our result, it is reasonable to neglect the contribution from the second-generation pairs. 2) We use the EBL spectrum and optical depth from the model C in Ref.~\citep{2010ApJ...712..238F} in the calculation. We check the results considering alternative EBL models from Ref.~\citep{2011MNRAS.410.2556D} and Ref.~\citep{2012MNRAS.422.3189G}, but find the difference is small.

In Ref.~\citep{2017ApJ...847...39V}, the authors used GRB 130427A  to constrain the IGMF by assuming that the GeV emission of GRB 130427A extends to $\sim 10$ TeV. They obtained a  limit of $>3\times 10^{-17}$ G for a coherence length of $\lambda = 1$ Mpc. Their limit is stronger than ours mainly because of two factors: 1) they miss the timescale $l_{\gamma\gamma}(\epsilon)\theta_{\rm B}^{2}(\gamma_{\rm e},B)/(2c)$ for the duration of the echo emission, which leads to an overestimate of the echo emission flux; 2) The assumed fluence in the TeV emission of GRB 130427A is  higher than that of GRB 190114C.

\begin{figure}[htbp]
\centering
\includegraphics[width=1\columnwidth]{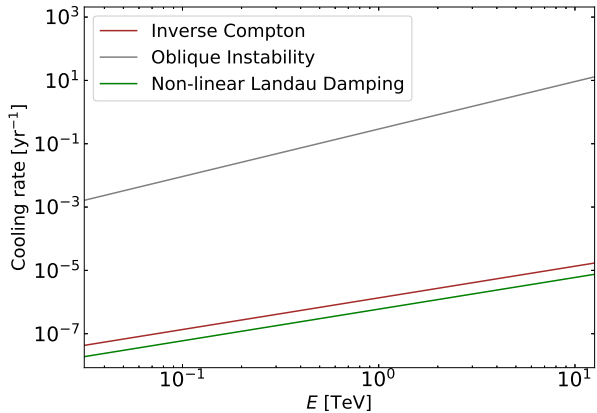}
\caption{Cooling rates ($-\dot{\gamma_{\rm e}}/\gamma_{\rm e}$) computed by using the parameters of GRB 190114C.
The red line represents the IC cooling rate. The gray liner line represents the cooling rate due to the oblique instability  \citep{2012ApJ...752...22B} and the gray liner line represents the cooling rate due to the non-linear Landau damping  \citep{2019MNRAS.489.3836A}.\label{fig:rate}}
\end{figure}
{In addition, the limit on  IGMF based on the cascade flux may become weaker or even avoided if the  plasma instabilities, arising due to the interaction of the electrons/positrons pairs with the intergalactic medium, cool down the pairs faster than the IC scattering, as has been discussed for the blazar case~\citep{2019MNRAS.489.3836A,2019arXiv190909210B,2019ApJ...870...17Y}. As an example, we study two kinds of plasma instability, i.e., the oblique instability and non-linear Landau damping, and then compare these  cooling rates with the IC cooling rate. The result is shown in Figure~\ref{fig:rate}. The oblique instability model is based on the result in ref.~\citep{2012ApJ...752...22B}, while the non-linear Landau damping model corresponds to model B in ref.~\citep{2019MNRAS.489.3836A}.
It can be seen from Figure~\ref{fig:rate} that the cooling rate for the non-linear Landau damping model is  lower than that of the IC process, so the  limit on IGMF remains almost unchanged. On the other hand,  for the oblique instability, the cooling rate is much higher than that of the IC process and hence no limit on IGMF can be obtained.  Other plasma instabilities may also occur \citep{2012ApJ...758..102S, 2013ApJ...777...49S, 2014ApJ...787...49S, 2018ApJ...857...43V, 2019MNRAS.489.3836A, 2019arXiv190909210B}, and the limit on the IGMF could be avoided for those with strong instability cooling. }

The limit on IGMF can be improved in future if  GRBs with larger fluences in TeV emission are detected \citep{2010Sci...328...73N}.
Another approach is to looking for the echo emission in the energy range of $>100$ GeV, since the duration becomes small and the flux increases. This may be achieved through observations of GRBs with higher sensitivity Cherenkov telescopes in the future, such as Cherenkov Telescope Array  \citep{2011ExA....32..193A}. However, there are some requirements for using these very high energy telescopes to limit IGMF: 1) As the echo photons with energy $E>100$ GeV are produced by the cascade process of high-energy photons with energy of $E\gtrsim 10$ TeV, this requires that GRBs can emit such high energy photons; 2) If the echo photon energy extends to the sub-TeV range, the multi-cascade process should then be considered.

{\em A note added: } {After the submission of our paper, a paper appears on arXiv \citep{2020arXiv200206918D}, arguing that the sensitivity of  Fermi-LAT  is not sufficient to obtain meaningful constraints on IGMF. However, in their calculation,  they only take into account the primary TeV photons during the period  from 62~s to 2454~s, neglecting the fact that the power-law decay of the afterglow flux starts from 6~s \citep{2019A&A...626A..12R, 2019ApJ...884..117W,2019Natur.575..459M}. This leads to that the energy of the primary TeV photons used in their calculation is about a factor of 5 lower than ours.}

We thank P. Veres for useful discussions. The work is supported by the NSFC under the grants 11625312 and 11851304, and the National Key
R\&D program of China under the grant 2018YFA0404203.


\begin{thebibliography}{47}%
\makeatletter
\providecommand \@ifxundefined [1]{%
 \@ifx{#1\undefined}
}%
\providecommand \@ifnum [1]{%
 \ifnum #1\expandafter \@firstoftwo
 \else \expandafter \@secondoftwo
 \fi
}%
\providecommand \@ifx [1]{%
 \ifx #1\expandafter \@firstoftwo
 \else \expandafter \@secondoftwo
 \fi
}%
\providecommand \natexlab [1]{#1}%
\providecommand \enquote  [1]{``#1''}%
\providecommand \bibnamefont  [1]{#1}%
\providecommand \bibfnamefont [1]{#1}%
\providecommand \citenamefont [1]{#1}%
\providecommand \href@noop [0]{\@secondoftwo}%
\providecommand \href [0]{\begingroup \@sanitize@url \@href}%
\providecommand \@href[1]{\@@startlink{#1}\@@href}%
\providecommand \@@href[1]{\endgroup#1\@@endlink}%
\providecommand \@sanitize@url [0]{\catcode `\\12\catcode `\$12\catcode
  `\&12\catcode `\#12\catcode `\^12\catcode `\_12\catcode `\%12\relax}%
\providecommand \@@startlink[1]{}%
\providecommand \@@endlink[0]{}%
\providecommand \url  [0]{\begingroup\@sanitize@url \@url }%
\providecommand \@url [1]{\endgroup\@href {#1}{\urlprefix }}%
\providecommand \urlprefix  [0]{URL }%
\providecommand \Eprint [0]{\href }%
\providecommand \doibase [0]{http://dx.doi.org/}%
\providecommand \selectlanguage [0]{\@gobble}%
\providecommand \bibinfo  [0]{\@secondoftwo}%
\providecommand \bibfield  [0]{\@secondoftwo}%
\providecommand \translation [1]{[#1]}%
\providecommand \BibitemOpen [0]{}%
\providecommand \bibitemStop [0]{}%
\providecommand \bibitemNoStop [0]{.\EOS\space}%
\providecommand \EOS [0]{\spacefactor3000\relax}%
\providecommand \BibitemShut  [1]{\csname bibitem#1\endcsname}%
\let\auto@bib@innerbib\@empty
\bibitem [{\citenamefont {{Durrer}}\ and\ \citenamefont
  {{Neronov}}(2013)}]{2013A&ARv..21...62D}%
  \BibitemOpen
  \bibfield  {author} {\bibinfo {author} {\bibfnamefont {R.}~\bibnamefont
  {{Durrer}}}\ and\ \bibinfo {author} {\bibfnamefont {A.}~\bibnamefont
  {{Neronov}}},\ }\href {\doibase 10.1007/s00159-013-0062-7} {\bibfield
  {journal} {\bibinfo  {journal} {\aap}\ }\textbf {\bibinfo {volume} {21}},\
  \bibinfo {eid} {62} (\bibinfo {year} {2013})},\ \Eprint
  {http://arxiv.org/abs/1303.7121} {arXiv:1303.7121 [astro-ph.CO]} \BibitemShut
  {NoStop}%
\bibitem [{\citenamefont {{Kulsrud}}\ and\ \citenamefont
  {{Zweibel}}(2008)}]{2008RPPh...71d6901K}%
  \BibitemOpen
  \bibfield  {author} {\bibinfo {author} {\bibfnamefont {R.~M.}\ \bibnamefont
  {{Kulsrud}}}\ and\ \bibinfo {author} {\bibfnamefont {E.~G.}\ \bibnamefont
  {{Zweibel}}},\ }\href {\doibase 10.1088/0034-4885/71/4/046901} {\bibfield
  {journal} {\bibinfo  {journal} {Reports on Progress in Physics}\ }\textbf
  {\bibinfo {volume} {71}},\ \bibinfo {eid} {046901} (\bibinfo {year}
  {2008})},\ \Eprint {http://arxiv.org/abs/0707.2783} {arXiv:0707.2783
  [astro-ph]} \BibitemShut {NoStop}%
\bibitem [{\citenamefont {{Neronov}}\ \emph {et~al.}(2013)\citenamefont
  {{Neronov}}, \citenamefont {{Taylor}}, \citenamefont {{Tchernin}},\ and\
  \citenamefont {{Vovk}}}]{2013A&A...554A..31N}%
  \BibitemOpen
  \bibfield  {author} {\bibinfo {author} {\bibfnamefont {A.}~\bibnamefont
  {{Neronov}}}, \bibinfo {author} {\bibfnamefont {A.~M.}\ \bibnamefont
  {{Taylor}}}, \bibinfo {author} {\bibfnamefont {C.}~\bibnamefont
  {{Tchernin}}}, \ and\ \bibinfo {author} {\bibfnamefont {I.}~\bibnamefont
  {{Vovk}}},\ }\href {\doibase 10.1051/0004-6361/201321294} {\bibfield
  {journal} {\bibinfo  {journal} {\aap}\ }\textbf {\bibinfo {volume} {554}},\
  \bibinfo {eid} {A31} (\bibinfo {year} {2013})},\ \Eprint
  {http://arxiv.org/abs/1307.2753} {arXiv:1307.2753 [astro-ph.HE]} \BibitemShut
  {NoStop}%
\bibitem [{\citenamefont {{Aharonian}}\ \emph {et~al.}(1994)\citenamefont
  {{Aharonian}}, \citenamefont {{Coppi}},\ and\ \citenamefont
  {{Voelk}}}]{1994ApJ...423L...5A}%
  \BibitemOpen
  \bibfield  {author} {\bibinfo {author} {\bibfnamefont {F.~A.}\ \bibnamefont
  {{Aharonian}}}, \bibinfo {author} {\bibfnamefont {P.~S.}\ \bibnamefont
  {{Coppi}}}, \ and\ \bibinfo {author} {\bibfnamefont {H.~J.}\ \bibnamefont
  {{Voelk}}},\ }\href {\doibase 10.1086/187222} {\bibfield  {journal} {\bibinfo
   {journal} {\apjl}\ }\textbf {\bibinfo {volume} {423}},\ \bibinfo {pages}
  {L5} (\bibinfo {year} {1994})},\ \Eprint
  {http://arxiv.org/abs/astro-ph/9312045} {arXiv:astro-ph/9312045 [astro-ph]}
  \BibitemShut {NoStop}%
\bibitem [{\citenamefont {{Tavecchio}}\ \emph {et~al.}(2010)\citenamefont
  {{Tavecchio}}, \citenamefont {{Ghisellini}}, \citenamefont {{Foschini}},
  \citenamefont {{Bonnoli}}, \citenamefont {{Ghirlanda}},\ and\ \citenamefont
  {{Coppi}}}]{2010MNRAS.406L..70T}%
  \BibitemOpen
  \bibfield  {author} {\bibinfo {author} {\bibfnamefont {F.}~\bibnamefont
  {{Tavecchio}}}, \bibinfo {author} {\bibfnamefont {G.}~\bibnamefont
  {{Ghisellini}}}, \bibinfo {author} {\bibfnamefont {L.}~\bibnamefont
  {{Foschini}}}, \bibinfo {author} {\bibfnamefont {G.}~\bibnamefont
  {{Bonnoli}}}, \bibinfo {author} {\bibfnamefont {G.}~\bibnamefont
  {{Ghirlanda}}}, \ and\ \bibinfo {author} {\bibfnamefont {P.}~\bibnamefont
  {{Coppi}}},\ }\href {\doibase 10.1111/j.1745-3933.2010.00884.x} {\bibfield
  {journal} {\bibinfo  {journal} {\mnras}\ }\textbf {\bibinfo {volume} {406}},\
  \bibinfo {pages} {L70} (\bibinfo {year} {2010})},\ \Eprint
  {http://arxiv.org/abs/1004.1329} {arXiv:1004.1329 [astro-ph.CO]} \BibitemShut
  {NoStop}%
\bibitem [{\citenamefont {{Tavecchio}}\ \emph {et~al.}(2011)\citenamefont
  {{Tavecchio}}, \citenamefont {{Ghisellini}}, \citenamefont {{Bonnoli}},\ and\
  \citenamefont {{Foschini}}}]{2011MNRAS.414.3566T}%
  \BibitemOpen
  \bibfield  {author} {\bibinfo {author} {\bibfnamefont {F.}~\bibnamefont
  {{Tavecchio}}}, \bibinfo {author} {\bibfnamefont {G.}~\bibnamefont
  {{Ghisellini}}}, \bibinfo {author} {\bibfnamefont {G.}~\bibnamefont
  {{Bonnoli}}}, \ and\ \bibinfo {author} {\bibfnamefont {L.}~\bibnamefont
  {{Foschini}}},\ }\href {\doibase 10.1111/j.1365-2966.2011.18657.x} {\bibfield
   {journal} {\bibinfo  {journal} {\mnras}\ }\textbf {\bibinfo {volume}
  {414}},\ \bibinfo {pages} {3566} (\bibinfo {year} {2011})},\ \Eprint
  {http://arxiv.org/abs/1009.1048} {arXiv:1009.1048 [astro-ph.HE]} \BibitemShut
  {NoStop}%
\bibitem [{\citenamefont {{Neronov}}\ \emph {et~al.}(2010)\citenamefont
  {{Neronov}}, \citenamefont {{Semikoz}}, \citenamefont {{Kachelriess}},
  \citenamefont {{Ostapchenko}},\ and\ \citenamefont
  {{Elyiv}}}]{2010ApJ...719L.130N}%
  \BibitemOpen
  \bibfield  {author} {\bibinfo {author} {\bibfnamefont {A.}~\bibnamefont
  {{Neronov}}}, \bibinfo {author} {\bibfnamefont {D.}~\bibnamefont
  {{Semikoz}}}, \bibinfo {author} {\bibfnamefont {M.}~\bibnamefont
  {{Kachelriess}}}, \bibinfo {author} {\bibfnamefont {S.}~\bibnamefont
  {{Ostapchenko}}}, \ and\ \bibinfo {author} {\bibfnamefont {A.}~\bibnamefont
  {{Elyiv}}},\ }\href {\doibase 10.1088/2041-8205/719/2/L130} {\bibfield
  {journal} {\bibinfo  {journal} {\apjl}\ }\textbf {\bibinfo {volume} {719}},\
  \bibinfo {pages} {L130} (\bibinfo {year} {2010})},\ \Eprint
  {http://arxiv.org/abs/1002.4981} {arXiv:1002.4981 [astro-ph.HE]} \BibitemShut
  {NoStop}%
\bibitem [{\citenamefont {{Dermer}}\ \emph {et~al.}(2011)\citenamefont
  {{Dermer}}, \citenamefont {{Cavadini}}, \citenamefont {{Razzaque}},
  \citenamefont {{Finke}}, \citenamefont {{Chiang}},\ and\ \citenamefont
  {{Lott}}}]{2011ApJ...733L..21D}%
  \BibitemOpen
  \bibfield  {author} {\bibinfo {author} {\bibfnamefont {C.~D.}\ \bibnamefont
  {{Dermer}}}, \bibinfo {author} {\bibfnamefont {M.}~\bibnamefont
  {{Cavadini}}}, \bibinfo {author} {\bibfnamefont {S.}~\bibnamefont
  {{Razzaque}}}, \bibinfo {author} {\bibfnamefont {J.~D.}\ \bibnamefont
  {{Finke}}}, \bibinfo {author} {\bibfnamefont {J.}~\bibnamefont {{Chiang}}}, \
  and\ \bibinfo {author} {\bibfnamefont {B.}~\bibnamefont {{Lott}}},\ }\href
  {\doibase 10.1088/2041-8205/733/2/L21} {\bibfield  {journal} {\bibinfo
  {journal} {\apjl}\ }\textbf {\bibinfo {volume} {733}},\ \bibinfo {pages}
  {L21} (\bibinfo {year} {2011})},\ \Eprint {http://arxiv.org/abs/1011.6660}
  {arXiv:1011.6660 [astro-ph.HE]} \BibitemShut {NoStop}%
\bibitem [{\citenamefont {{Arlen}}\ \emph {et~al.}(2014)\citenamefont
  {{Arlen}}, \citenamefont {{Vassilev}}, \citenamefont {{Weisgarber}},
  \citenamefont {{Wakely}},\ and\ \citenamefont {{Yusef
  Shafi}}}]{2014ApJ...796...18A}%
  \BibitemOpen
  \bibfield  {author} {\bibinfo {author} {\bibfnamefont {T.~C.}\ \bibnamefont
  {{Arlen}}}, \bibinfo {author} {\bibfnamefont {V.~V.}\ \bibnamefont
  {{Vassilev}}}, \bibinfo {author} {\bibfnamefont {T.}~\bibnamefont
  {{Weisgarber}}}, \bibinfo {author} {\bibfnamefont {S.~P.}\ \bibnamefont
  {{Wakely}}}, \ and\ \bibinfo {author} {\bibfnamefont {S.}~\bibnamefont
  {{Yusef Shafi}}},\ }\href {\doibase 10.1088/0004-637X/796/1/18} {\bibfield
  {journal} {\bibinfo  {journal} {\apj}\ }\textbf {\bibinfo {volume} {796}},\
  \bibinfo {eid} {18} (\bibinfo {year} {2014})},\ \Eprint
  {http://arxiv.org/abs/1210.2802} {arXiv:1210.2802 [astro-ph.HE]} \BibitemShut
  {NoStop}%
\bibitem [{\citenamefont {{Finke}}\ \emph {et~al.}(2015)\citenamefont
  {{Finke}}, \citenamefont {{Reyes}}, \citenamefont {{Georganopoulos}},
  \citenamefont {{Reynolds}}, \citenamefont {{Ajello}}, \citenamefont
  {{Fegan}},\ and\ \citenamefont {{McCann}}}]{2015ApJ...814...20F}%
  \BibitemOpen
  \bibfield  {author} {\bibinfo {author} {\bibfnamefont {J.~D.}\ \bibnamefont
  {{Finke}}}, \bibinfo {author} {\bibfnamefont {L.~C.}\ \bibnamefont
  {{Reyes}}}, \bibinfo {author} {\bibfnamefont {M.}~\bibnamefont
  {{Georganopoulos}}}, \bibinfo {author} {\bibfnamefont {K.}~\bibnamefont
  {{Reynolds}}}, \bibinfo {author} {\bibfnamefont {M.}~\bibnamefont
  {{Ajello}}}, \bibinfo {author} {\bibfnamefont {S.~J.}\ \bibnamefont
  {{Fegan}}}, \ and\ \bibinfo {author} {\bibfnamefont {K.}~\bibnamefont
  {{McCann}}},\ }\href {\doibase 10.1088/0004-637X/814/1/20} {\bibfield
  {journal} {\bibinfo  {journal} {\apj}\ }\textbf {\bibinfo {volume} {814}},\
  \bibinfo {eid} {20} (\bibinfo {year} {2015})},\ \Eprint
  {http://arxiv.org/abs/1510.02485} {arXiv:1510.02485 [astro-ph.HE]}
  \BibitemShut {NoStop}%
\bibitem [{\citenamefont {{D'Avezac}}\ \emph {et~al.}(2007)\citenamefont
  {{D'Avezac}}, \citenamefont {{Dubus}},\ and\ \citenamefont
  {{Giebels}}}]{2007A&A...469..857D}%
  \BibitemOpen
  \bibfield  {author} {\bibinfo {author} {\bibfnamefont {P.}~\bibnamefont
  {{D'Avezac}}}, \bibinfo {author} {\bibfnamefont {G.}~\bibnamefont {{Dubus}}},
  \ and\ \bibinfo {author} {\bibfnamefont {B.}~\bibnamefont {{Giebels}}},\
  }\href {\doibase 10.1051/0004-6361:20066712} {\bibfield  {journal} {\bibinfo
  {journal} {\aap}\ }\textbf {\bibinfo {volume} {469}},\ \bibinfo {pages} {857}
  (\bibinfo {year} {2007})},\ \Eprint {http://arxiv.org/abs/0704.3910}
  {arXiv:0704.3910 [astro-ph]} \BibitemShut {NoStop}%
\bibitem [{\citenamefont {{Neronov}}\ and\ \citenamefont
  {{Vovk}}(2010)}]{2010Sci...328...73N}%
  \BibitemOpen
  \bibfield  {author} {\bibinfo {author} {\bibfnamefont {A.}~\bibnamefont
  {{Neronov}}}\ and\ \bibinfo {author} {\bibfnamefont {I.}~\bibnamefont
  {{Vovk}}},\ }\href {\doibase 10.1126/science.1184192} {\bibfield  {journal}
  {\bibinfo  {journal} {Science}\ }\textbf {\bibinfo {volume} {328}},\ \bibinfo
  {pages} {73} (\bibinfo {year} {2010})},\ \Eprint
  {http://arxiv.org/abs/1006.3504} {arXiv:1006.3504 [astro-ph.HE]} \BibitemShut
  {NoStop}%
\bibitem [{\citenamefont {{Ackermann}}\ \emph {et~al.}(2018)\citenamefont
  {{Ackermann}}, \citenamefont {{Ajello}}, \citenamefont {{Baldini}},
  \citenamefont {{Ballet}}, \citenamefont {{Barbiellini}}, \citenamefont
  {{Bastieri}}, \citenamefont {{Bellazzini}}, \citenamefont {{Bissaldi}},
  \citenamefont {{Blandford}}, \citenamefont {{Bloom}},\ and\ \citenamefont
  {et~al.}}]{2018ApJS..237...32A}%
  \BibitemOpen
  \bibfield  {author} {\bibinfo {author} {\bibfnamefont {M.}~\bibnamefont
  {{Ackermann}}}, \bibinfo {author} {\bibfnamefont {M.}~\bibnamefont
  {{Ajello}}}, \bibinfo {author} {\bibfnamefont {L.}~\bibnamefont {{Baldini}}},
  \bibinfo {author} {\bibfnamefont {J.}~\bibnamefont {{Ballet}}}, \bibinfo
  {author} {\bibfnamefont {G.}~\bibnamefont {{Barbiellini}}}, \bibinfo {author}
  {\bibfnamefont {D.}~\bibnamefont {{Bastieri}}}, \bibinfo {author}
  {\bibfnamefont {R.}~\bibnamefont {{Bellazzini}}}, \bibinfo {author}
  {\bibfnamefont {E.}~\bibnamefont {{Bissaldi}}}, \bibinfo {author}
  {\bibfnamefont {R.~D.}\ \bibnamefont {{Blandford}}}, \bibinfo {author}
  {\bibfnamefont {E.~D.}\ \bibnamefont {{Bloom}}}, \ and\ \bibinfo {author}
  {\bibnamefont {et~al.}},\ }\href {\doibase 10.3847/1538-4365/aacdf7}
  {\bibfield  {journal} {\bibinfo  {journal} {\apjs}\ }\textbf {\bibinfo
  {volume} {237}},\ \bibinfo {eid} {32} (\bibinfo {year} {2018})},\ \Eprint
  {http://arxiv.org/abs/1804.08035} {arXiv:1804.08035 [astro-ph.HE]}
  \BibitemShut {NoStop}%
\bibitem [{\citenamefont {{Plaga}}(1995)}]{1995Natur.374..430P}%
  \BibitemOpen
  \bibfield  {author} {\bibinfo {author} {\bibfnamefont {R.}~\bibnamefont
  {{Plaga}}},\ }\href {\doibase 10.1038/374430a0} {\bibfield  {journal}
  {\bibinfo  {journal} {\nat}\ }\textbf {\bibinfo {volume} {374}},\ \bibinfo
  {pages} {430} (\bibinfo {year} {1995})}\BibitemShut {NoStop}%
\bibitem [{\citenamefont {{Dai}}\ \emph {et~al.}(2002)\citenamefont {{Dai}},
  \citenamefont {{Zhang}}, \citenamefont {{Gou}}, \citenamefont
  {{M{\'e}sz{\'a}ros}},\ and\ \citenamefont {{Waxman}}}]{2002ApJ...580L...7D}%
  \BibitemOpen
  \bibfield  {author} {\bibinfo {author} {\bibfnamefont {Z.~G.}\ \bibnamefont
  {{Dai}}}, \bibinfo {author} {\bibfnamefont {B.}~\bibnamefont {{Zhang}}},
  \bibinfo {author} {\bibfnamefont {L.~J.}\ \bibnamefont {{Gou}}}, \bibinfo
  {author} {\bibfnamefont {P.}~\bibnamefont {{M{\'e}sz{\'a}ros}}}, \ and\
  \bibinfo {author} {\bibfnamefont {E.}~\bibnamefont {{Waxman}}},\ }\href
  {\doibase 10.1086/345494} {\bibfield  {journal} {\bibinfo  {journal} {\apjl}\
  }\textbf {\bibinfo {volume} {580}},\ \bibinfo {pages} {L7} (\bibinfo {year}
  {2002})},\ \Eprint {http://arxiv.org/abs/astro-ph/0209091}
  {arXiv:astro-ph/0209091 [astro-ph]} \BibitemShut {NoStop}%
\bibitem [{\citenamefont {{Fan}}\ \emph {et~al.}(2004)\citenamefont {{Fan}},
  \citenamefont {{Dai}},\ and\ \citenamefont {{Wei}}}]{2004A&A...415..483F}%
  \BibitemOpen
  \bibfield  {author} {\bibinfo {author} {\bibfnamefont {Y.~Z.}\ \bibnamefont
  {{Fan}}}, \bibinfo {author} {\bibfnamefont {Z.~G.}\ \bibnamefont {{Dai}}}, \
  and\ \bibinfo {author} {\bibfnamefont {D.~M.}\ \bibnamefont {{Wei}}},\ }\href
  {\doibase 10.1051/0004-6361:20034472} {\bibfield  {journal} {\bibinfo
  {journal} {\aap}\ }\textbf {\bibinfo {volume} {415}},\ \bibinfo {pages} {483}
  (\bibinfo {year} {2004})},\ \Eprint {http://arxiv.org/abs/astro-ph/0310893}
  {arXiv:astro-ph/0310893 [astro-ph]} \BibitemShut {NoStop}%
\bibitem [{\citenamefont {{Murase}}\ \emph {et~al.}(2008)\citenamefont
  {{Murase}}, \citenamefont {{Takahashi}}, \citenamefont {{Inoue}},
  \citenamefont {{Ichiki}},\ and\ \citenamefont
  {{Nagataki}}}]{2008ApJ...686L..67M}%
  \BibitemOpen
  \bibfield  {author} {\bibinfo {author} {\bibfnamefont {K.}~\bibnamefont
  {{Murase}}}, \bibinfo {author} {\bibfnamefont {K.}~\bibnamefont
  {{Takahashi}}}, \bibinfo {author} {\bibfnamefont {S.}~\bibnamefont
  {{Inoue}}}, \bibinfo {author} {\bibfnamefont {K.}~\bibnamefont {{Ichiki}}}, \
  and\ \bibinfo {author} {\bibfnamefont {S.}~\bibnamefont {{Nagataki}}},\
  }\href {\doibase 10.1086/592997} {\bibfield  {journal} {\bibinfo  {journal}
  {\apjl}\ }\textbf {\bibinfo {volume} {686}},\ \bibinfo {pages} {L67}
  (\bibinfo {year} {2008})},\ \Eprint {http://arxiv.org/abs/0806.2829}
  {arXiv:0806.2829 [astro-ph]} \BibitemShut {NoStop}%
\bibitem [{\citenamefont {{Takahashi}}\ \emph {et~al.}(2012)\citenamefont
  {{Takahashi}}, \citenamefont {{Mori}}, \citenamefont {{Ichiki}},\ and\
  \citenamefont {{Inoue}}}]{2012ApJ...744L...7T}%
  \BibitemOpen
  \bibfield  {author} {\bibinfo {author} {\bibfnamefont {K.}~\bibnamefont
  {{Takahashi}}}, \bibinfo {author} {\bibfnamefont {M.}~\bibnamefont {{Mori}}},
  \bibinfo {author} {\bibfnamefont {K.}~\bibnamefont {{Ichiki}}}, \ and\
  \bibinfo {author} {\bibfnamefont {S.}~\bibnamefont {{Inoue}}},\ }\href
  {\doibase 10.1088/2041-8205/744/1/L7} {\bibfield  {journal} {\bibinfo
  {journal} {\apjl}\ }\textbf {\bibinfo {volume} {744}},\ \bibinfo {eid} {L7}
  (\bibinfo {year} {2012})},\ \Eprint {http://arxiv.org/abs/1103.3835}
  {arXiv:1103.3835 [astro-ph.CO]} \BibitemShut {NoStop}%
\bibitem [{\citenamefont {{Wang}}\ \emph {et~al.}(2004)\citenamefont {{Wang}},
  \citenamefont {{Cheng}}, \citenamefont {{Dai}},\ and\ \citenamefont
  {{Lu}}}]{2004ApJ...604..306W}%
  \BibitemOpen
  \bibfield  {author} {\bibinfo {author} {\bibfnamefont {X.~Y.}\ \bibnamefont
  {{Wang}}}, \bibinfo {author} {\bibfnamefont {K.~S.}\ \bibnamefont {{Cheng}}},
  \bibinfo {author} {\bibfnamefont {Z.~G.}\ \bibnamefont {{Dai}}}, \ and\
  \bibinfo {author} {\bibfnamefont {T.}~\bibnamefont {{Lu}}},\ }\href {\doibase
  10.1086/381745} {\bibfield  {journal} {\bibinfo  {journal} {\apj}\ }\textbf
  {\bibinfo {volume} {604}},\ \bibinfo {pages} {306} (\bibinfo {year}
  {2004})},\ \Eprint {http://arxiv.org/abs/astro-ph/0311601}
  {arXiv:astro-ph/0311601 [astro-ph]} \BibitemShut {NoStop}%
\bibitem [{\citenamefont {{Razzaque}}\ \emph {et~al.}(2004)\citenamefont
  {{Razzaque}}, \citenamefont {{M{\'e}sz{\'a}ros}},\ and\ \citenamefont
  {{Zhang}}}]{2004ApJ...613.1072R}%
  \BibitemOpen
  \bibfield  {author} {\bibinfo {author} {\bibfnamefont {S.}~\bibnamefont
  {{Razzaque}}}, \bibinfo {author} {\bibfnamefont {P.}~\bibnamefont
  {{M{\'e}sz{\'a}ros}}}, \ and\ \bibinfo {author} {\bibfnamefont
  {B.}~\bibnamefont {{Zhang}}},\ }\href {\doibase 10.1086/423166} {\bibfield
  {journal} {\bibinfo  {journal} {\apj}\ }\textbf {\bibinfo {volume} {613}},\
  \bibinfo {pages} {1072} (\bibinfo {year} {2004})},\ \Eprint
  {http://arxiv.org/abs/astro-ph/0404076} {arXiv:astro-ph/0404076 [astro-ph]}
  \BibitemShut {NoStop}%
\bibitem [{\citenamefont {{Ichiki}}\ \emph {et~al.}(2008)\citenamefont
  {{Ichiki}}, \citenamefont {{Inoue}},\ and\ \citenamefont
  {{Takahashi}}}]{2008ApJ...682..127I}%
  \BibitemOpen
  \bibfield  {author} {\bibinfo {author} {\bibfnamefont {K.}~\bibnamefont
  {{Ichiki}}}, \bibinfo {author} {\bibfnamefont {S.}~\bibnamefont {{Inoue}}}, \
  and\ \bibinfo {author} {\bibfnamefont {K.}~\bibnamefont {{Takahashi}}},\
  }\href {\doibase 10.1086/588275} {\bibfield  {journal} {\bibinfo  {journal}
  {\apj}\ }\textbf {\bibinfo {volume} {682}},\ \bibinfo {pages} {127} (\bibinfo
  {year} {2008})},\ \Eprint {http://arxiv.org/abs/0711.1589} {arXiv:0711.1589
  [astro-ph]} \BibitemShut {NoStop}%
\bibitem [{\citenamefont {{Murase}}\ \emph {et~al.}(2009)\citenamefont
  {{Murase}}, \citenamefont {{Zhang}}, \citenamefont {{Takahashi}},\ and\
  \citenamefont {{Nagataki}}}]{2009MNRAS.396.1825M}%
  \BibitemOpen
  \bibfield  {author} {\bibinfo {author} {\bibfnamefont {K.}~\bibnamefont
  {{Murase}}}, \bibinfo {author} {\bibfnamefont {B.}~\bibnamefont {{Zhang}}},
  \bibinfo {author} {\bibfnamefont {K.}~\bibnamefont {{Takahashi}}}, \ and\
  \bibinfo {author} {\bibfnamefont {S.}~\bibnamefont {{Nagataki}}},\ }\href
  {\doibase 10.1111/j.1365-2966.2009.14704.x} {\bibfield  {journal} {\bibinfo
  {journal} {\mnras}\ }\textbf {\bibinfo {volume} {396}},\ \bibinfo {pages}
  {1825} (\bibinfo {year} {2009})},\ \Eprint {http://arxiv.org/abs/0812.0124}
  {arXiv:0812.0124 [astro-ph]} \BibitemShut {NoStop}%
\bibitem [{\citenamefont {{Takahashi}}\ \emph {et~al.}(2011)\citenamefont
  {{Takahashi}}, \citenamefont {{Inoue}}, \citenamefont {{Ichiki}},\ and\
  \citenamefont {{Nakamura}}}]{2011MNRAS.410.2741T}%
  \BibitemOpen
  \bibfield  {author} {\bibinfo {author} {\bibfnamefont {K.}~\bibnamefont
  {{Takahashi}}}, \bibinfo {author} {\bibfnamefont {S.}~\bibnamefont
  {{Inoue}}}, \bibinfo {author} {\bibfnamefont {K.}~\bibnamefont {{Ichiki}}}, \
  and\ \bibinfo {author} {\bibfnamefont {T.}~\bibnamefont {{Nakamura}}},\
  }\href {\doibase 10.1111/j.1365-2966.2010.17639.x} {\bibfield  {journal}
  {\bibinfo  {journal} {\mnras}\ }\textbf {\bibinfo {volume} {410}},\ \bibinfo
  {pages} {2741} (\bibinfo {year} {2011})},\ \Eprint
  {http://arxiv.org/abs/1007.5363} {arXiv:1007.5363 [astro-ph.HE]} \BibitemShut
  {NoStop}%
\bibitem [{\citenamefont {{Veres}}\ \emph {et~al.}(2017)\citenamefont
  {{Veres}}, \citenamefont {{Dermer}},\ and\ \citenamefont
  {{Dhuga}}}]{2017ApJ...847...39V}%
  \BibitemOpen
  \bibfield  {author} {\bibinfo {author} {\bibfnamefont {P.}~\bibnamefont
  {{Veres}}}, \bibinfo {author} {\bibfnamefont {C.~D.}\ \bibnamefont
  {{Dermer}}}, \ and\ \bibinfo {author} {\bibfnamefont {K.~S.}\ \bibnamefont
  {{Dhuga}}},\ }\href {\doibase 10.3847/1538-4357/aa87b1} {\bibfield  {journal}
  {\bibinfo  {journal} {\apj}\ }\textbf {\bibinfo {volume} {847}},\ \bibinfo
  {eid} {39} (\bibinfo {year} {2017})},\ \Eprint
  {http://arxiv.org/abs/1705.08531} {arXiv:1705.08531 [astro-ph.HE]}
  \BibitemShut {NoStop}%
\bibitem [{\citenamefont {{MAGIC Collaboration}}\ \emph
  {et~al.}(2019{\natexlab{a}})\citenamefont {{MAGIC Collaboration}},
  \citenamefont {{Acciari}}, \citenamefont {{Ansoldi}}, \citenamefont
  {{Antonelli}}, \citenamefont {{Arbet Engels}}, \citenamefont {{Baack}},
  \citenamefont {{Babi{\'c}}}, \citenamefont {{Banerjee}}, \citenamefont
  {{Barres de Almeida}}, \citenamefont {{Barrio}},\ and\ \citenamefont
  {et~al.}}]{2019Natur.575..455M}%
  \BibitemOpen
  \bibfield  {author} {\bibinfo {author} {\bibnamefont {{MAGIC
  Collaboration}}}, \bibinfo {author} {\bibfnamefont {V.~A.}\ \bibnamefont
  {{Acciari}}}, \bibinfo {author} {\bibfnamefont {S.}~\bibnamefont
  {{Ansoldi}}}, \bibinfo {author} {\bibfnamefont {L.~A.}\ \bibnamefont
  {{Antonelli}}}, \bibinfo {author} {\bibfnamefont {A.}~\bibnamefont {{Arbet
  Engels}}}, \bibinfo {author} {\bibfnamefont {D.}~\bibnamefont {{Baack}}},
  \bibinfo {author} {\bibfnamefont {A.}~\bibnamefont {{Babi{\'c}}}}, \bibinfo
  {author} {\bibfnamefont {B.}~\bibnamefont {{Banerjee}}}, \bibinfo {author}
  {\bibfnamefont {U.}~\bibnamefont {{Barres de Almeida}}}, \bibinfo {author}
  {\bibfnamefont {J.~A.}\ \bibnamefont {{Barrio}}}, \ and\ \bibinfo {author}
  {\bibnamefont {et~al.}},\ }\href {\doibase 10.1038/s41586-019-1750-x}
  {\bibfield  {journal} {\bibinfo  {journal} {\nat}\ }\textbf {\bibinfo
  {volume} {575}},\ \bibinfo {pages} {455} (\bibinfo {year}
  {2019}{\natexlab{a}})}\BibitemShut {NoStop}%
\bibitem [{\citenamefont {{MAGIC Collaboration}}\ \emph
  {et~al.}(2019{\natexlab{b}})\citenamefont {{MAGIC Collaboration}},
  \citenamefont {{Acciari}}, \citenamefont {{Ansoldi}}, \citenamefont
  {{Antonelli}}, \citenamefont {{Engels}}, \citenamefont {{Baack}},
  \citenamefont {{Babi{\'c}}}, \citenamefont {{Banerjee}}, \citenamefont
  {{Barres de Almeida}}, \citenamefont {{Barrio}},\ and\ \citenamefont
  {et~al.}}]{2019Natur.575..459M}%
  \BibitemOpen
  \bibfield  {author} {\bibinfo {author} {\bibnamefont {{MAGIC
  Collaboration}}}, \bibinfo {author} {\bibfnamefont {V.~A.}\ \bibnamefont
  {{Acciari}}}, \bibinfo {author} {\bibfnamefont {S.}~\bibnamefont
  {{Ansoldi}}}, \bibinfo {author} {\bibfnamefont {L.~A.}\ \bibnamefont
  {{Antonelli}}}, \bibinfo {author} {\bibfnamefont {A.~A.}\ \bibnamefont
  {{Engels}}}, \bibinfo {author} {\bibfnamefont {D.}~\bibnamefont {{Baack}}},
  \bibinfo {author} {\bibfnamefont {A.}~\bibnamefont {{Babi{\'c}}}}, \bibinfo
  {author} {\bibfnamefont {B.}~\bibnamefont {{Banerjee}}}, \bibinfo {author}
  {\bibfnamefont {U.}~\bibnamefont {{Barres de Almeida}}}, \bibinfo {author}
  {\bibfnamefont {J.~A.}\ \bibnamefont {{Barrio}}}, \ and\ \bibinfo {author}
  {\bibnamefont {et~al.}},\ }\href {\doibase 10.1038/s41586-019-1754-6}
  {\bibfield  {journal} {\bibinfo  {journal} {\nat}\ }\textbf {\bibinfo
  {volume} {575}},\ \bibinfo {pages} {459} (\bibinfo {year}
  {2019}{\natexlab{b}})}\BibitemShut {NoStop}%
\bibitem [{\citenamefont {{Selsing}}\ \emph {et~al.}(2019)\citenamefont
  {{Selsing}}, \citenamefont {{Fynbo}}, \citenamefont {{Heintz}},\ and\
  \citenamefont {{Watson}}}]{2019GCN.23695....1S}%
  \BibitemOpen
  \bibfield  {author} {\bibinfo {author} {\bibfnamefont {J.}~\bibnamefont
  {{Selsing}}}, \bibinfo {author} {\bibfnamefont {J.~P.~U.}\ \bibnamefont
  {{Fynbo}}}, \bibinfo {author} {\bibfnamefont {K.~E.}\ \bibnamefont
  {{Heintz}}}, \ and\ \bibinfo {author} {\bibfnamefont {D.}~\bibnamefont
  {{Watson}}},\ }\href@noop {} {\bibfield  {journal} {\bibinfo  {journal} {GRB
  Coordinates Network}\ }\textbf {\bibinfo {volume} {23695}},\ \bibinfo {pages}
  {1} (\bibinfo {year} {2019})}\BibitemShut {NoStop}%
\bibitem [{\citenamefont {{Zdziarski}}(1988)}]{1988ApJ...335..786Z}%
  \BibitemOpen
  \bibfield  {author} {\bibinfo {author} {\bibfnamefont {A.~A.}\ \bibnamefont
  {{Zdziarski}}},\ }\href {\doibase 10.1086/166967} {\bibfield  {journal}
  {\bibinfo  {journal} {\apj}\ }\textbf {\bibinfo {volume} {335}},\ \bibinfo
  {pages} {786} (\bibinfo {year} {1988})}\BibitemShut {NoStop}%
\bibitem [{\citenamefont {{Jones}}(1968)}]{1968PhRv..167.1159J}%
  \BibitemOpen
  \bibfield  {author} {\bibinfo {author} {\bibfnamefont {F.~C.}\ \bibnamefont
  {{Jones}}},\ }\href {\doibase 10.1103/PhysRev.167.1159} {\bibfield  {journal}
  {\bibinfo  {journal} {Physical Review}\ }\textbf {\bibinfo {volume} {167}},\
  \bibinfo {pages} {1159} (\bibinfo {year} {1968})}\BibitemShut {NoStop}%
\bibitem [{\citenamefont {{Ravasio}}\ \emph {et~al.}(2019)\citenamefont
  {{Ravasio}}, \citenamefont {{Oganesyan}}, \citenamefont {{Salafia}},
  \citenamefont {{Ghirland a}}, \citenamefont {{Ghisellini}}, \citenamefont
  {{Branchesi}}, \citenamefont {{Campana}}, \citenamefont {{Covino}},\ and\
  \citenamefont {{Salvaterra}}}]{2019A&A...626A..12R}%
  \BibitemOpen
  \bibfield  {author} {\bibinfo {author} {\bibfnamefont {M.~E.}\ \bibnamefont
  {{Ravasio}}}, \bibinfo {author} {\bibfnamefont {G.}~\bibnamefont
  {{Oganesyan}}}, \bibinfo {author} {\bibfnamefont {O.~S.}\ \bibnamefont
  {{Salafia}}}, \bibinfo {author} {\bibfnamefont {G.}~\bibnamefont {{Ghirland
  a}}}, \bibinfo {author} {\bibfnamefont {G.}~\bibnamefont {{Ghisellini}}},
  \bibinfo {author} {\bibfnamefont {M.}~\bibnamefont {{Branchesi}}}, \bibinfo
  {author} {\bibfnamefont {S.}~\bibnamefont {{Campana}}}, \bibinfo {author}
  {\bibfnamefont {S.}~\bibnamefont {{Covino}}}, \ and\ \bibinfo {author}
  {\bibfnamefont {R.}~\bibnamefont {{Salvaterra}}},\ }\href {\doibase
  10.1051/0004-6361/201935214} {\bibfield  {journal} {\bibinfo  {journal}
  {\aap}\ }\textbf {\bibinfo {volume} {626}},\ \bibinfo {eid} {A12} (\bibinfo
  {year} {2019})},\ \Eprint {http://arxiv.org/abs/1902.01861} {arXiv:1902.01861
  [astro-ph.HE]} \BibitemShut {NoStop}%
\bibitem [{\citenamefont {{Wang}}\ \emph {et~al.}(2019)\citenamefont {{Wang}},
  \citenamefont {{Liu}}, \citenamefont {{Zhang}}, \citenamefont {{Xi}},\ and\
  \citenamefont {{Zhang}}}]{2019ApJ...884..117W}%
  \BibitemOpen
  \bibfield  {author} {\bibinfo {author} {\bibfnamefont {X.-Y.}\ \bibnamefont
  {{Wang}}}, \bibinfo {author} {\bibfnamefont {R.-Y.}\ \bibnamefont {{Liu}}},
  \bibinfo {author} {\bibfnamefont {H.-M.}\ \bibnamefont {{Zhang}}}, \bibinfo
  {author} {\bibfnamefont {S.-Q.}\ \bibnamefont {{Xi}}}, \ and\ \bibinfo
  {author} {\bibfnamefont {B.}~\bibnamefont {{Zhang}}},\ }\href {\doibase
  10.3847/1538-4357/ab426c} {\bibfield  {journal} {\bibinfo  {journal} {\apj}\
  }\textbf {\bibinfo {volume} {884}},\ \bibinfo {eid} {117} (\bibinfo {year}
  {2019})},\ \Eprint {http://arxiv.org/abs/1905.11312} {arXiv:1905.11312
  [astro-ph.HE]} \BibitemShut {NoStop}%
\bibitem [{\citenamefont {{The Fermi-LAT
  collaboration}}(2019)}]{2019arXiv190210045T}%
  \BibitemOpen
  \bibfield  {author} {\bibinfo {author} {\bibnamefont {{The Fermi-LAT
  collaboration}}},\ }\href@noop {} {\bibfield  {journal} {\bibinfo  {journal}
  {arXiv e-prints}\ ,\ \bibinfo {eid} {arXiv:1902.10045}} (\bibinfo {year}
  {2019})},\ \Eprint {http://arxiv.org/abs/1902.10045} {arXiv:1902.10045
  [astro-ph.HE]} \BibitemShut {NoStop}%
\bibitem [{Note1()}]{Note1}%
  \BibitemOpen
  \bibinfo {note} {\protect \textit {gll\protect \_iem\protect \_v07.fits} and
  \protect \textit {iso\protect \_P8R3\protect \_SOURCE\protect \_V2\protect
  \_1.txt};
  https://fermi.gsfc.nasa.gov/ssc/data/access/lat/BackgroundModels.html}\BibitemShut
  {NoStop}%
\bibitem [{Note2()}]{Note2}%
  \BibitemOpen
  \bibinfo {note} {Https://gcn.gsfc.nasa.gov/gcn3/23724.gcn3}\BibitemShut
  {NoStop}%
\bibitem [{\citenamefont {{Finke}}\ \emph {et~al.}(2010)\citenamefont
  {{Finke}}, \citenamefont {{Razzaque}},\ and\ \citenamefont
  {{Dermer}}}]{2010ApJ...712..238F}%
  \BibitemOpen
  \bibfield  {author} {\bibinfo {author} {\bibfnamefont {J.~D.}\ \bibnamefont
  {{Finke}}}, \bibinfo {author} {\bibfnamefont {S.}~\bibnamefont {{Razzaque}}},
  \ and\ \bibinfo {author} {\bibfnamefont {C.~D.}\ \bibnamefont {{Dermer}}},\
  }\href {\doibase 10.1088/0004-637X/712/1/238} {\bibfield  {journal} {\bibinfo
   {journal} {\apj}\ }\textbf {\bibinfo {volume} {712}},\ \bibinfo {pages}
  {238} (\bibinfo {year} {2010})},\ \Eprint {http://arxiv.org/abs/0905.1115}
  {arXiv:0905.1115 [astro-ph.HE]} \BibitemShut {NoStop}%
\bibitem [{\citenamefont {{Dom{\'\i}nguez}}\ \emph {et~al.}(2011)\citenamefont
  {{Dom{\'\i}nguez}}, \citenamefont {{Primack}}, \citenamefont {{Rosario}},
  \citenamefont {{Prada}}, \citenamefont {{Gilmore}}, \citenamefont {{Faber}},
  \citenamefont {{Koo}}, \citenamefont {{Somerville}}, \citenamefont
  {{P{\'e}rez-Torres}}, \citenamefont {{P{\'e}rez-Gonz{\'a}lez}},\ and\
  \citenamefont {et~al.}}]{2011MNRAS.410.2556D}%
  \BibitemOpen
  \bibfield  {author} {\bibinfo {author} {\bibfnamefont {A.}~\bibnamefont
  {{Dom{\'\i}nguez}}}, \bibinfo {author} {\bibfnamefont {J.~R.}\ \bibnamefont
  {{Primack}}}, \bibinfo {author} {\bibfnamefont {D.~J.}\ \bibnamefont
  {{Rosario}}}, \bibinfo {author} {\bibfnamefont {F.}~\bibnamefont {{Prada}}},
  \bibinfo {author} {\bibfnamefont {R.~C.}\ \bibnamefont {{Gilmore}}}, \bibinfo
  {author} {\bibfnamefont {S.~M.}\ \bibnamefont {{Faber}}}, \bibinfo {author}
  {\bibfnamefont {D.~C.}\ \bibnamefont {{Koo}}}, \bibinfo {author}
  {\bibfnamefont {R.~S.}\ \bibnamefont {{Somerville}}}, \bibinfo {author}
  {\bibfnamefont {M.~A.}\ \bibnamefont {{P{\'e}rez-Torres}}}, \bibinfo {author}
  {\bibfnamefont {P.}~\bibnamefont {{P{\'e}rez-Gonz{\'a}lez}}}, \ and\ \bibinfo
  {author} {\bibnamefont {et~al.}},\ }\href {\doibase
  10.1111/j.1365-2966.2010.17631.x} {\bibfield  {journal} {\bibinfo  {journal}
  {\mnras}\ }\textbf {\bibinfo {volume} {410}},\ \bibinfo {pages} {2556}
  (\bibinfo {year} {2011})},\ \Eprint {http://arxiv.org/abs/1007.1459}
  {arXiv:1007.1459 [astro-ph.CO]} \BibitemShut {NoStop}%
\bibitem [{\citenamefont {{Gilmore}}\ \emph {et~al.}(2012)\citenamefont
  {{Gilmore}}, \citenamefont {{Somerville}}, \citenamefont {{Primack}},\ and\
  \citenamefont {{Dom{\'\i}nguez}}}]{2012MNRAS.422.3189G}%
  \BibitemOpen
  \bibfield  {author} {\bibinfo {author} {\bibfnamefont {R.~C.}\ \bibnamefont
  {{Gilmore}}}, \bibinfo {author} {\bibfnamefont {R.~S.}\ \bibnamefont
  {{Somerville}}}, \bibinfo {author} {\bibfnamefont {J.~R.}\ \bibnamefont
  {{Primack}}}, \ and\ \bibinfo {author} {\bibfnamefont {A.}~\bibnamefont
  {{Dom{\'\i}nguez}}},\ }\href {\doibase 10.1111/j.1365-2966.2012.20841.x}
  {\bibfield  {journal} {\bibinfo  {journal} {\mnras}\ }\textbf {\bibinfo
  {volume} {422}},\ \bibinfo {pages} {3189} (\bibinfo {year} {2012})},\ \Eprint
  {http://arxiv.org/abs/1104.0671} {arXiv:1104.0671 [astro-ph.CO]} \BibitemShut
  {NoStop}%
\bibitem [{\citenamefont {{Broderick}}\ \emph {et~al.}(2012)\citenamefont
  {{Broderick}}, \citenamefont {{Chang}},\ and\ \citenamefont
  {{Pfrommer}}}]{2012ApJ...752...22B}%
  \BibitemOpen
  \bibfield  {author} {\bibinfo {author} {\bibfnamefont {A.~E.}\ \bibnamefont
  {{Broderick}}}, \bibinfo {author} {\bibfnamefont {P.}~\bibnamefont
  {{Chang}}}, \ and\ \bibinfo {author} {\bibfnamefont {C.}~\bibnamefont
  {{Pfrommer}}},\ }\href {\doibase 10.1088/0004-637X/752/1/22} {\bibfield
  {journal} {\bibinfo  {journal} {\apj}\ }\textbf {\bibinfo {volume} {752}},\
  \bibinfo {eid} {22} (\bibinfo {year} {2012})},\ \Eprint
  {http://arxiv.org/abs/1106.5494} {arXiv:1106.5494 [astro-ph.CO]} \BibitemShut
  {NoStop}%
\bibitem [{\citenamefont {{Alves Batista}}\ \emph {et~al.}(2019)\citenamefont
  {{Alves Batista}}, \citenamefont {{Saveliev}},\ and\ \citenamefont {{de
  Gouveia Dal Pino}}}]{2019MNRAS.489.3836A}%
  \BibitemOpen
  \bibfield  {author} {\bibinfo {author} {\bibfnamefont {R.}~\bibnamefont
  {{Alves Batista}}}, \bibinfo {author} {\bibfnamefont {A.}~\bibnamefont
  {{Saveliev}}}, \ and\ \bibinfo {author} {\bibfnamefont {E.~M.}\ \bibnamefont
  {{de Gouveia Dal Pino}}},\ }\href {\doibase 10.1093/mnras/stz2389} {\bibfield
   {journal} {\bibinfo  {journal} {\mnras}\ }\textbf {\bibinfo {volume}
  {489}},\ \bibinfo {pages} {3836} (\bibinfo {year} {2019})},\ \Eprint
  {http://arxiv.org/abs/1904.13345} {arXiv:1904.13345 [astro-ph.HE]}
  \BibitemShut {NoStop}%
\bibitem [{\citenamefont {{Yan}}\ \emph {et~al.}(2019)\citenamefont {{Yan}},
  \citenamefont {{Zhou}}, \citenamefont {{Zhang}}, \citenamefont {{Zhu}},\ and\
  \citenamefont {{Wang}}}]{2019ApJ...870...17Y}%
  \BibitemOpen
  \bibfield  {author} {\bibinfo {author} {\bibfnamefont {D.}~\bibnamefont
  {{Yan}}}, \bibinfo {author} {\bibfnamefont {J.}~\bibnamefont {{Zhou}}},
  \bibinfo {author} {\bibfnamefont {P.}~\bibnamefont {{Zhang}}}, \bibinfo
  {author} {\bibfnamefont {Q.}~\bibnamefont {{Zhu}}}, \ and\ \bibinfo {author}
  {\bibfnamefont {J.}~\bibnamefont {{Wang}}},\ }\href {\doibase
  10.3847/1538-4357/aaef7d} {\bibfield  {journal} {\bibinfo  {journal} {\apj}\
  }\textbf {\bibinfo {volume} {870}},\ \bibinfo {eid} {17} (\bibinfo {year}
  {2019})},\ \Eprint {http://arxiv.org/abs/1810.07013} {arXiv:1810.07013
  [astro-ph.HE]} \BibitemShut {NoStop}%
\bibitem [{\citenamefont {{Blytt}}\ \emph {et~al.}(2019)\citenamefont
  {{Blytt}}, \citenamefont {{Kachelriess}},\ and\ \citenamefont
  {{Ostapchenko}}}]{2019arXiv190909210B}%
  \BibitemOpen
  \bibfield  {author} {\bibinfo {author} {\bibfnamefont {M.}~\bibnamefont
  {{Blytt}}}, \bibinfo {author} {\bibfnamefont {M.}~\bibnamefont
  {{Kachelriess}}}, \ and\ \bibinfo {author} {\bibfnamefont {S.}~\bibnamefont
  {{Ostapchenko}}},\ }\href@noop {} {\bibfield  {journal} {\bibinfo  {journal}
  {arXiv e-prints}\ ,\ \bibinfo {eid} {arXiv:1909.09210}} (\bibinfo {year}
  {2019})},\ \Eprint {http://arxiv.org/abs/1909.09210} {arXiv:1909.09210
  [astro-ph.HE]} \BibitemShut {NoStop}%
\bibitem [{\citenamefont {{Schlickeiser}}\ \emph {et~al.}(2012)\citenamefont
  {{Schlickeiser}}, \citenamefont {{Ibscher}},\ and\ \citenamefont
  {{Supsar}}}]{2012ApJ...758..102S}%
  \BibitemOpen
  \bibfield  {author} {\bibinfo {author} {\bibfnamefont {R.}~\bibnamefont
  {{Schlickeiser}}}, \bibinfo {author} {\bibfnamefont {D.}~\bibnamefont
  {{Ibscher}}}, \ and\ \bibinfo {author} {\bibfnamefont {M.}~\bibnamefont
  {{Supsar}}},\ }\href {\doibase 10.1088/0004-637X/758/2/102} {\bibfield
  {journal} {\bibinfo  {journal} {\apj}\ }\textbf {\bibinfo {volume} {758}},\
  \bibinfo {eid} {102} (\bibinfo {year} {2012})}\BibitemShut {NoStop}%
\bibitem [{\citenamefont {{Schlickeiser}}\ \emph {et~al.}(2013)\citenamefont
  {{Schlickeiser}}, \citenamefont {{Krakau}},\ and\ \citenamefont
  {{Supsar}}}]{2013ApJ...777...49S}%
  \BibitemOpen
  \bibfield  {author} {\bibinfo {author} {\bibfnamefont {R.}~\bibnamefont
  {{Schlickeiser}}}, \bibinfo {author} {\bibfnamefont {S.}~\bibnamefont
  {{Krakau}}}, \ and\ \bibinfo {author} {\bibfnamefont {M.}~\bibnamefont
  {{Supsar}}},\ }\href {\doibase 10.1088/0004-637X/777/1/49} {\bibfield
  {journal} {\bibinfo  {journal} {\apj}\ }\textbf {\bibinfo {volume} {777}},\
  \bibinfo {eid} {49} (\bibinfo {year} {2013})},\ \Eprint
  {http://arxiv.org/abs/1308.4594} {arXiv:1308.4594 [astro-ph.HE]} \BibitemShut
  {NoStop}%
\bibitem [{\citenamefont {{Sironi}}\ and\ \citenamefont
  {{Giannios}}(2014)}]{2014ApJ...787...49S}%
  \BibitemOpen
  \bibfield  {author} {\bibinfo {author} {\bibfnamefont {L.}~\bibnamefont
  {{Sironi}}}\ and\ \bibinfo {author} {\bibfnamefont {D.}~\bibnamefont
  {{Giannios}}},\ }\href {\doibase 10.1088/0004-637X/787/1/49} {\bibfield
  {journal} {\bibinfo  {journal} {\apj}\ }\textbf {\bibinfo {volume} {787}},\
  \bibinfo {eid} {49} (\bibinfo {year} {2014})},\ \Eprint
  {http://arxiv.org/abs/1312.4538} {arXiv:1312.4538 [astro-ph.HE]} \BibitemShut
  {NoStop}%
\bibitem [{\citenamefont {{Vafin}}\ \emph {et~al.}(2018)\citenamefont
  {{Vafin}}, \citenamefont {{Rafighi}}, \citenamefont {{Pohl}},\ and\
  \citenamefont {{Niemiec}}}]{2018ApJ...857...43V}%
  \BibitemOpen
  \bibfield  {author} {\bibinfo {author} {\bibfnamefont {S.}~\bibnamefont
  {{Vafin}}}, \bibinfo {author} {\bibfnamefont {I.}~\bibnamefont {{Rafighi}}},
  \bibinfo {author} {\bibfnamefont {M.}~\bibnamefont {{Pohl}}}, \ and\ \bibinfo
  {author} {\bibfnamefont {J.}~\bibnamefont {{Niemiec}}},\ }\href {\doibase
  10.3847/1538-4357/aab552} {\bibfield  {journal} {\bibinfo  {journal} {\apj}\
  }\textbf {\bibinfo {volume} {857}},\ \bibinfo {eid} {43} (\bibinfo {year}
  {2018})},\ \Eprint {http://arxiv.org/abs/1803.02990} {arXiv:1803.02990
  [astro-ph.HE]} \BibitemShut {NoStop}%
\bibitem [{\citenamefont {{Actis}}\ \emph {et~al.}(2011)\citenamefont
  {{Actis}}, \citenamefont {{Agnetta}}, \citenamefont {{Aharonian}},
  \citenamefont {{Akhperjanian}}, \citenamefont {{Aleksi{\'c}}}, \citenamefont
  {{Aliu}}, \citenamefont {{Allan}}, \citenamefont {{Allekotte}}, \citenamefont
  {{Antico}}, \citenamefont {{Antonelli}},\ and\ \citenamefont
  {et~al.}}]{2011ExA....32..193A}%
  \BibitemOpen
  \bibfield  {author} {\bibinfo {author} {\bibfnamefont {M.}~\bibnamefont
  {{Actis}}}, \bibinfo {author} {\bibfnamefont {G.}~\bibnamefont {{Agnetta}}},
  \bibinfo {author} {\bibfnamefont {F.}~\bibnamefont {{Aharonian}}}, \bibinfo
  {author} {\bibfnamefont {A.}~\bibnamefont {{Akhperjanian}}}, \bibinfo
  {author} {\bibfnamefont {J.}~\bibnamefont {{Aleksi{\'c}}}}, \bibinfo {author}
  {\bibfnamefont {E.}~\bibnamefont {{Aliu}}}, \bibinfo {author} {\bibfnamefont
  {D.}~\bibnamefont {{Allan}}}, \bibinfo {author} {\bibfnamefont
  {I.}~\bibnamefont {{Allekotte}}}, \bibinfo {author} {\bibfnamefont
  {F.}~\bibnamefont {{Antico}}}, \bibinfo {author} {\bibfnamefont {L.~A.}\
  \bibnamefont {{Antonelli}}}, \ and\ \bibinfo {author} {\bibnamefont
  {et~al.}},\ }\href {\doibase 10.1007/s10686-011-9247-0} {\bibfield  {journal}
  {\bibinfo  {journal} {Experimental Astronomy}\ }\textbf {\bibinfo {volume}
  {32}},\ \bibinfo {pages} {193} (\bibinfo {year} {2011})},\ \Eprint
  {http://arxiv.org/abs/1008.3703} {arXiv:1008.3703 [astro-ph.IM]} \BibitemShut
  {NoStop}%
\bibitem [{\citenamefont {{Dzhatdoev}}\ and\ \citenamefont
  {{Podlesnyi}}(2020)}]{2020arXiv200206918D}%
  \BibitemOpen
  \bibfield  {author} {\bibinfo {author} {\bibfnamefont {T.~A.}\ \bibnamefont
  {{Dzhatdoev}}}\ and\ \bibinfo {author} {\bibfnamefont {E.~I.}\ \bibnamefont
  {{Podlesnyi}}},\ }\href@noop {} {\bibfield  {journal} {\bibinfo  {journal}
  {arXiv e-prints}\ ,\ \bibinfo {eid} {arXiv:2002.06918}} (\bibinfo {year}
  {2020})},\ \Eprint {http://arxiv.org/abs/2002.06918} {arXiv:2002.06918
  [astro-ph.HE]} \BibitemShut {NoStop}%
\end{thebibliography}
%

\end{document}